\documentstyle[a4dutch]{article}
\begin {document}
\thispagestyle {empty}
\begin{flushright}
Preprint: arch-ive/9609229
\end{flushright}
\begin{center}
\vspace {5mm}
{\large\bf MINIMAL SPINNING STRING\\}
\vspace {7mm}
Yu.R.~Musin\\
\vspace {4mm}
{\it Department of Applied Mathematics, Moscow Aviation
Institute,\\
Volokolamskoe shosse 4, Moscow, 125871, Russia\\}
\vspace{4mm}
{\tt musin@k804.mainet.msk.su\\}
\vspace {5mm}
\end {center}

\begin{abstract}
{\rm
Minimal $N=1/2$ supersymmetric extension of bosonic Polyakov's string
are constructed. This model is natural generalization of Di~Vecchia-Ravndal
superparticle. The classical sector of the model are investigated, Noether
currents and Virosoro supercondition are found. Minimal spinning string is
more simple, than the standard $N=1$ spinning string of Neveu-Schwarz-Ramond
and has a number of unusial properties such as a chiral symmetry, parabolic
type of equations of movement, non-triviality fermionic sectors for
closed strings only and e.t.c.}
\end{abstract}

Standard $N=1$ supersymmetric extension bosonic strings admits, as is known,
only two variants of further expansion -- $N=2$ and $N=4$\,\cite{theorss}.
Unexplored, however, there is the sector of the half-whole dimensions, the
existence of which can be assumed, proceeding from existence of
the appropriate
theories of superparticles. A minimal superparticle of such type
is $N=1/2$ a
superparticle Di~Vecchia-Ravndal\,\cite{divrav, musin}. The purpose of the
given work is construction $N=1/2$ of supersymmetric expansion
bosonic strings.
Such spin string is minimal string model, from which superparticle
Di~Vecchia-Ravndal can be received at aspiration of the size of a string to
zero.

We shall consider in $D$-dimensional Minkowski space  a world sheet,
describing bosonic a string $x^{\mu} (\tau, \,\sigma),\: \mu =
0,\,1,\,2,\,
\ldots ,\,D-1$, where $\tau$ and $\sigma$ -- according to
temporary and space parameter of points on sheet.

We shall generalize space of parameters $\sigma^a =
\{\tau,\,\sigma\}, \
a=1,\,2$ up to superspace $B_L^{2,\,1} = \{\sigma^a,\,\theta\} =\mathstrut^0
B_L \otimes \mathstrut^0 B_L \otimes \mathstrut^1 B_L$, where $\mathstrut^0
B_L, \, \mathstrut^1 B_L$ -- accordingly an even and odd subset of Grassmann
algebra  $B_L$\,\cite {freund}. On superspace $B^{2,\,1}_L$ there are the
transformations of parameters, realizing the group SUSY:
\begin{equation}
\label{susy}
\begin{array}{cc}
\medskip
Translation\:(T): & Supertransformation\:(S):\\
\left\{
\begin{array}{llll}
\sigma^a & \rightarrow & \sigma^a + \alpha^a\\
\theta & \rightarrow & \theta &  \alpha\in\mathstrut^0 B_L\\
\end{array}
\right.
&
\left\{
\begin{array}{llll}
\sigma^a & \rightarrow & \sigma^a + i\rho^a\varepsilon\theta\\
\theta & \rightarrow & \theta + \varepsilon &
\varepsilon\in\mathstrut^1 B_L\:.\\
\end{array}
\right.
\end{array}
\end{equation}
Here $\rho^a$ -- some numerical constants. At transformations $T$ and
$S$ (\ref
{susy}) is saved 1-form
\begin{equation}
\label{form}
\omega^a = d \sigma^a - i \rho^a\theta d \theta\:,
\end{equation}
specifying metric in superspace
\begin{equation}
\label{metric}
d S^2 = \eta_{ab} \omega^a\omega_b\:.
\end{equation}
The metric tensor can be chosen in conformal-flat form $\eta_{ab} =
{\rm diag}
(-1,\,1)$\,\cite{theorss}.

Thus on $B^{2,1}_L$ superexpansion of Minkowski metric is received
\begin{equation}
\label{metric2}
\begin{array}{c}
\medskip
d S^2 = \eta_{ab} \omega^a\omega^b = g_{KL} d \Omega^K d \Omega^L
\:; \qquad K,\, L = 0,\,1,\,2\\
\begin{array}{cc}
\left(
\begin{array}{ccc}
-1 & 0 & i\rho^0\theta\\
0 & -1 & -i\rho^1\theta\\
i\rho^0\theta & -i\rho^1\theta & 0\\
\end{array}
\right) \quad
d \Omega^K = \{d\tau, \, d\sigma, \, d\theta\}
\end{array}
\end{array}
\end{equation}

On superspace of parameters $B^{2,\,1}_L$ it is possible to set superfield
\begin {equation}
\label {sfield}
X^{\mu}: B^{2,\,1}_L \longrightarrow \bigotimes\limits^ {D-1}_{\mu = 0}
\mathstrut^0 B_L \;, \end {equation}
which final decomposition by virtue of nilpotence $\theta\;
(\theta^2 = 0)$ has
\begin{equation}
\label{expand}
X^{\mu} (\tau,\,\sigma,\,\theta) = x^{\mu} (\tau,\,\sigma) + i \psi^{\mu}
(\tau,\,\sigma) \theta\;, \quad x^{\mu} \in \mathstrut^0 B_L \,,\;\psi^{\mu}
\in \mathstrut^1 B_L \,.
\end {equation}
We shall identify $x^{\mu} (\tau,\,\sigma)$ with coordinates of points of a
string in $D$-dimensional Minkowski space, and $\psi^{\mu} (\tau,\,\sigma)$
with $D$-multiplet majoran fermionic of fields, transforming on vector
presentation of Lorentz group $SO (D-1,\,1)$.

The group SUSY has linear representation on vector superfields $X^{\mu}$
over $B^{2,\,1}_L$. The generators of this representation are easily
written out:
\begin{equation}
\label{gener}
\hat{P} _a = \left. {\frac{dT [X]}{d\alpha^a}} \right |_{\alpha = 0} =
\frac {\partial} {\partial\sigma^a} \,;\quad \hat {Q} = \left.\frac {dS [X]}
{d\varepsilon} \right | _ {\varepsilon = 0} = \frac {\vec {\partial}}
{\partial\theta} + i \rho^a\theta \frac {\partial} {\partial\sigma^a} \;.
\end{equation}
Covariant derivative operator
\begin{equation}
\label{covar}
D = \frac {\vec {\partial}} {\partial\theta} - i \rho^a \theta \partial_a;
\quad \partial_a = \frac {\partial} {\partial \sigma^a} \end {equation}
together with generators $\hat {P} _a$ and $\hat {Q} $ derivate
algebra SUSY:
\begin{equation}
\label{algebra}
\begin{array}{c}
\left [\hat {P} _a, \, \hat {P} _b \right] = \left[\hat {P} _a, \, \hat {Q}
\right] = \left[D, \, \hat {P} _a \right] = 0 \;; \nonumber\\
\left\{\hat {Q}, \, D
\right\} = 0 \;; \quad \left\{\hat {Q}, \, \hat {Q} \right\} = - \Bigl\{D, \,
D \Bigr\} = 2 i \rho^a \hat {P} _a \,.
\end{array}
\end{equation}
The formulas of transformation of components fields $x^ {\mu} $ and $\psi^
{\mu} $ are easily written out. So at supertransformation $S$ we have
\begin{equation}
\label{variat}
\delta x^ {\mu} = - i \varepsilon \psi^ {\mu}; \; \delta \psi^ {\mu} =
\varepsilon \rho^a \partial_a x^ {\mu} \;.
\end{equation}

We shall proceed now to construction of dynamics of a string. We shall posit
action, similar to action for bosonic Polyakov's string  in conformal
gauge\,\cite{kaf}
\begin{equation}
\label{action}
S = - \frac {1} {2\pi} \int d^2 \sigma \partial_a x^ {\mu} \partial^a x_
{\mu}
\end{equation}
and passing in action for superparticle
Di~Vecchia-Ravndal\,\cite{divrav}
\begin{equation}
\label {action2}
S = \frac {1} {2} \int d\tau \int d\theta \bar {D} X^ {\mu} \bar {D} ^2 X_
{\mu} \,,
\end{equation}
where $\bar {D} = \frac {\vec {\partial}} {\partial\theta} - i
\theta \frac {\partial} {\partial\tau} $.

Minimal action, generalizing (\ref {action}) for type (\ref
{action2}) -- it is obvious, that
\begin {equation}
\label {action3}
S = - \frac {1} {2\pi} \int d^2 \sigma \int d \theta D X^ {\mu} D^2 X_ {\mu}
\,.
\end {equation}
The lagrangian, adequate such action, can be written out in kind
\begin{equation}
\label{lagr}
L = - \frac {1} {2\pi} \int d\theta DX^ {\mu} D^2 X_ {\mu} = \frac {1} {2\pi}
\left (\rho^a\rho^b\partial_a x^ {\mu} \partial_b x_ {\mu} + i \rho^a \psi^
{\mu} \partial_a \psi_ {\mu} \right) \,. \end {equation}
For any action of a type (\ref {action3})
\begin {equation}
\label {action4}
S = \int d^2 \sigma L (\partial x, \, \partial \psi,\, \psi) \, \end
{equation}
from variational principle $\delta S = 0$ equations of a movement
can be received
\begin{equation}
\label{motion}
\frac {\partial} {\partial \tau} \frac {\partial L} {\partial \dot {x}} +
\frac
{\partial} {\partial\sigma} \frac {\partial L} {\partial x^ {\prime}} = 0 \,,
\quad \frac {\partial L} {\partial \psi} \frac {\partial} {\partial\tau}
\frac
{\partial L} {\partial \dot {\psi}} - \frac {\partial} {\partial\sigma}
\frac
{\partial L} {\partial \psi^ {\prime}} = 0 \,,
\end{equation}
where $\dot {A} \equiv \frac {\partial A^ {\mu}} {\partial\tau},
\, A^ {\prime}
\equiv \frac {\partial A^ {\mu}} {\partial\sigma} $ and boundary conditions
(for open string)
\begin{equation}
\label{bound}
\frac {\partial L} {\partial x^\prime} \delta x = 0 \,; \quad \frac {\partial
L} {\partial \psi^\prime} \delta \psi = 0 \quad at\ \sigma = 0, \, \pi\,.
\end{equation}

For closed string the boundary conditions (\ref {bound}) are away, instead of
them we shall impose a condition of periodicity
\begin{equation}
\label{period}
X^ {\mu} (\tau,\, 0) = x^ {\mu} (\tau,\, \pi); \quad \psi^ {\mu} (\tau,\, 0)
=
\pm \psi^ {\mu} (\tau,\, \pi) \,.
\end{equation}
By virtue of quadric of observable sizes on $\psi^ {\mu} $ the conditions
(\ref
{period}) do not result in discontinuity of observable values.

Substituting in (\ref {motion}) lagrangian (\ref {lagr}), we shall receive
equations of a movement
\begin{equation}
\label{motion2}
\rho^a \rho^b \partial^2_ {ab} x^ {\mu} = 0 \,; \quad \rho^a \partial_a \psi^
{\mu} = 0 \,,
\end{equation}
which can be rewritten in form
\begin{equation}
\label{motion3}
\ddot {x} + \alpha^2 x^ {\prime\prime} + 2 \alpha \dot {x} ^ {\prime} \,;
\quad
\dot {\psi} + \alpha^ {\prime} = 0 \,,
\end{equation}
where an vectorial index $\mu$ is omitted and
designation $\alpha \equiv \rho^1 / \rho^0$ is entered.

The solution (\ref {motion3}) has a kind
\begin{equation}
\label{sol}
x = \phi (\sigma - \alpha \tau) + \Phi (\sigma - \alpha \tau) \tau \,;
\quad
\psi = \psi (\sigma - \alpha \tau) \,,
\end{equation}
where $\phi, \, \Phi, \, \psi$ -- any smooth functions. Boundary
conditions (\ref {bound}) bring in requirement
\begin{equation}
\label{cond}
\Phi \equiv 0 \,; \quad \psi \equiv 0 \,,
\end{equation}
that is, despite parabolic type of the first of equations (\ref
{motion3}), it
has the solution as constant wave. The second of conditions (\ref
{cond}) means, that the open minimal string can be only bosonic.

For closed string of a condition of periodicity (\ref {period}) bring in
requirement of periodicity of functions $\phi, \, \Phi, \, \psi$ on $\tau$
with
period $\pi\alpha^{-1}$.

For finding-out of physical sense of the solution (\ref {sol}) we
shall find
canonical density of a momentum for (\ref {lagr})
\begin{equation}
\label{momentum}
p^a = - \frac {\partial L} {\partial\partial_a x} = - \frac {1} {\pi} \rho^a
\rho^b \partial_b x \,.
\end{equation}
Substituting here decision (\ref{sol}), we shall receive
\begin{equation}
\label{momcomp}
p^0 = - \frac {1} {\pi} \left (\rho^0 \right)^2 \Phi \,, \quad p^1 =
-\frac{1}{\pi} \rho^0 \rho^1 \Phi \,.
\end{equation}
The momentum of the whole string is received by integration
\begin{equation}
\label{momtot}
P = \int_0^ {\pi} p^0 d \sigma = - \frac {1} {\pi} \left (\rho^0 \right) ^2
\int_0^{\pi} \Phi (\sigma - \alpha\tau) d \sigma \,.
\end{equation}

Hence, for open string $P = 0$. For closed string we shall choose the even
solution $\Phi = - \left(\rho^0 \right)^{-2} \ P = const$, which
answers a
movement of a string as whole with momentum $P$. Then the solution
(\ref{sol}) can be rewritten in kind
\begin{equation}
\label{serie}
X^ {\mu} (\tau,\, \sigma) = x^{\mu} (0) + P^{\mu} \tau + \sum_ {n \ne 0}
\frac {a_n^ {\mu}} {n} e^ {-2 i n (\alpha\tau \sigma)} \,.
\end{equation}
Is here restored $D$-dimensional an index $\mu$ and the function
$\phi^{\mu}
(\sigma - \alpha\tau)$ is spreaded out on waves, running left to right at
$\alpha > 0$ ($R$-mode) and to the left at $\alpha < 0$
($L$-mode).

The fermionic sector of a closed string as against open
non-trivial and
consists of four parts: $L$-$R$, $L$-$NS$, $R$-$R$, $R$-$NS$. The second
symbol
means a type of periodic conditions (\ref{period})
\begin{equation}
\label{period2}
\begin{array}{c}
\medskip
\psi (\tau,\, 0) = \psi (\tau,\, \pi) - Ramond\ (R) \,, \nonumber\\
\medskip
\psi (\tau,\, 0)
= -\psi (\tau,\, \pi) - Neveu-Schwarz\ (NS)\,,\nonumber\\
\medskip
\psi^ {\mu} (\tau,\, \sigma) = \sum_ {n \in K} b_n e^ {-2 i n (\alpha\tau -
\sigma)} \,,\\
K = \left\{
\begin{array}{ll}
Z & Ramond\ L-R\ (\alpha < 0)\ and\ R-R\ (\alpha > 0)\\
Z + \frac{1}{2} & Neveu-Schwarz\ L-NS\ (\alpha < 0)\ and\
R-NS\ (\alpha > 0)\,.
\end{array}
\right.
\nonumber
\end{array}
\end{equation}
For research of constructed model we shall find Noether
currents~\cite{theorss}.

>From Lorentz-invariancy of lagrangian (\ref{lagr}) law of preservation
follows
\begin{equation}
\label{curr}
\partial_a J_ {\mu\nu} ^a = 0 \,,
\end{equation}
Where $J_ {\mu\nu} ^a = - \frac {\rho_a} {\pi} [\rho^b (x^ {\mu} \partial_b
x^
{\nu} - x^ {\nu} \partial_b x^ {\mu}) + i \psi^ {\mu} \psi^ {\nu}]$ - density
of tensor of a angular momentum. This law provides preservation
of a angular momentum
of a string, calculated on any spatial-like curve,
crossing an once a
world surface of a string.

Noether currents, caused by invariancy (\ref{lagr}) concerning
Poincar\'e{}-translations -- this bosonic $p_{\mu} ^a$ and
fermionic $\pi_{\mu} ^a$ of density of a momentum
\begin{equation}
\label{densmom}
p_{\mu} ^a = - \frac {1} {\pi} \rho^a \rho^b \partial_b x_ {\mu};
\quad
\pi_{\mu} ^a = \frac {i} {2 \pi} \rho^a \psi_ {\mu} \,.
\end{equation}
The laws of preservation -- it is simple equations of a movement
(\ref{motion2}).

By virtue of invariancy (\ref {lagr}) concerning sheet translations $T$
(\ref
{susy}) we have the law of preservation of tensor of an
energy-momentum
\begin{equation}
\label{cons}
\begin{array}{c}
\medskip
\partial_b T_b^a = 0 \,,\\
2 \pi T_b^a = 2 \rho^a \rho^c \partial_c x^ {\mu} \partial_b x_ {\mu} + i
\rho^a \psi^ {\mu} \partial_b \psi_ {\mu} - \delta_b^a \left (\rho^c \rho^d
\partial_c x^ {\mu} \partial_d x_ {\mu} + i \rho^c \psi^ {\mu} \partial_c
\psi_{\mu} \right) \,.
\end{array}
\end{equation}

And, at last, by virtue of invariancy (\ref {lagr}) concerning
supertransformation of components fields (\ref {variat}) we have
a preserved supercurrent $J^a$
\begin{equation}
\label{conscurr}
\partial_a J^a = 0 \,, \quad 2 \pi J^a = - i \rho^a \rho^b \partial_b x^{\mu}
\psi_{\mu} \,.
\end{equation}

As well as at all string models in theory of a minimal string
there are the
constraints. So from (\ref {densmom}) follows, that bosonic and
fermionic
density of a momentum of a string submit to constraints
\begin{equation}
\label{link}
P_1^{\mu} = \alpha P_0^ {\mu} \,, \quad \pi_1^ {\mu} = \alpha \pi_0^ {\mu}
\,.
\end{equation}
The choice $\alpha$ can be concretized, if to notice, that $\rho^a \rho_a =
\left (\rho^0 \right) ^2 (\alpha^2 - 1) $. Assuming $\rho^a \rho_a \equiv 0$,
we have $\alpha = \pm 1$, and, hence, the components of a supercurrent will
satisfy in this case to an additional  consractions
\begin {equation}
\label {link2}
\rho^a J_a = 0 \,.
\end {equation}
Easily to see, that by virtue of equations of a movement (\ref {motion3}) the
tensor of an energy-moment is traceless, as well as in bosonic case at $N =
0$
\begin{equation}
\label{spur}
Tr\,T_{ab} = 0 \,.
\end{equation}
Counting $T_{ab}$ symmetric, we shall receive a condition
\begin{equation}
\label{cond2}
\left(\dot {x} + \alpha x^ {\prime} \right) ^2 = P^ {\mu} P_ {\mu} = 0 \,,
\end{equation}
that is string as whole moves with velocity of light.

By analogy to coordinates of a light cone in purely bosonic case we shall
enter
semicone  coordinates
\begin {equation}
\label {cone}
t = \tau \,, \quad \xi = \sigma - \alpha \tau \,. \end {equation}
Then in these coordinates tensor of an energy-momentum of signs
a kind
\begin {equation}
\label {enmomcone}
2 \pi T_ {ab} = \left (
\begin {array} {cc}
T_ {tt} & T_ {t\xi} \\
T_ {\xi t} & T_ {\xi\xi}
\end {array}
\right) = - \left (\rho^0 \right) ^2 \left (
\begin {array} {cc}
- 2 (\dot {x} + \alpha x^ {\prime}) ^2 &
( \dot {x} + \alpha x^ {\prime}) ^2 \\
( \dot {x} + \alpha x^ {\prime}) ^2 &
\dot {x} ^2 - x^ {\prime 2} + \frac {i} {\rho^0} \psi \dot {\psi} \end
{array}
\right) \,,
\end {equation}
and vector of a supercurrent will change
\begin {equation}
\label {curcone}
J_t = 0 \,, \quad J_ {\xi} = - \frac {i} {2 \pi} \rho^0 \rho^1 \left (\dot {x}
+ \alpha x^ {\prime} \right) \psi \,. \end {equation}
The constraints then can be compactly written in form
\begin{equation}
\label{link3}
T_{tt} = 0 \,, \quad T_ {\xi\xi} = 0 \,, \quad J_ {\xi} = 0
\end{equation}
or in components
\begin {equation}
\label {link4}
\left (\dot {x} + \alpha x^ {\prime} \right) ^2 = 0 \,; \quad \dot {x} ^2 - x^
{\prime 2} + \frac {i} {\rho^0} \psi \dot {\psi} = 0 \,; \quad \left (\dot {x}
+ \alpha x^ {\prime} \right) \psi = 0 \,.
\end {equation}
The condition (\ref {link3}) or (\ref {link4}) can be posited as Virasoro
superconditions. It is desirable, certainly, to have a conclusion of these
conditions from more regular procedure, connected to fixing of calibration in
lagrangian of a string, constructed for type of a 2-dimensional supergravity.
Such problem is much difficult and represents the following stage of
development of the theory. The Virasoro conditions  are more important for
construction of the quantum theory of a minimal string, that is
problem, which is not mentioned in given work.

>From constructed classical theory of a minimal string it is visible, that the
received design is excellent from Neveu-Schwarz-Ramond string: asymmetry right
and left-hand; non-triviality fermionic sectors only for closed strings; a
movement of the whole string as whole with velocity of light; a
parabolic type of
equations of a movement and e.t.. As the theory of a minimal
string in some
sense is more simple, than the standard theory  of Neveu-Schwarz-Ramond, are
necessary further researches of this model, especially in quantum area.

\end {document}